# Labeling and Retrieval of Emotionally-Annotated Images using WordNet

Marko Horvat[†], Anton Grbin, Gordan Gledec

University of Zagreb, Faculty of Electrical Engineering and Computing,
Unska 3, HR-10000 Zagreb, Croatia
`{Marko.Horvat2, Anton.Grbin, Gordan.Gledec}@fer.hr`

**Abstract.** Repositories of images with semantic and emotion content descriptions are valuable tools in many areas such as Affective Computing and Human-Computer Interaction, but they are also important in the development of multimodal searchable online databases. Ever growing number of image documents available on the Internet continuously motivates research of better annotation models and more efficient retrieval methods which use mash-up of available data on semantics, scenes, objects, events, context and emotion. Formal knowledge representation of such high-level semantics requires rich, explicit, human but also machine-processable information. To achieve these goals we present an online ontology-based image annotation tool WNtags and demonstrate its usefulness in knowledge representation and image retrieval using the International Affective Picture System database. The WNtags uses WordNet as image tagging glossary but considers Suggested Upper Merged Ontology as the preferred upper labeling formalism. The retrieval is performed using node distance metrics to establish semantic relatedness between a query and the collaboratively weighted tags describing high-level image semantics, after which the result is ranked according to the derived importance. We also elaborate plans to improve the WNtags to create a collaborative Web-based multimedia repository for research in human emotion and attention.

## 1 Introduction

Prevalence of visual media on the Web emphasizes practical importance of image repositories that facilitate image storage and retrieval. Image retrieval algorithms ensuring high precision and recall, low fall-out rate and human and machine understandable vocabularies are being increasingly sought as the number of pictures available on the Internet is increases. Image semantics may be automatically recognized using image processing content detectors or manually set by human experts; either way the creation of a quality dataset of annotated images is an exhaustive and costly endeavor. Therefore, it becomes important to create a flexible and modular system that reuses available off-the-shelve systems (such as the tagging glossary or semantic relatedness module) as much as possible. The "WordNet tags" (or just "WNtags" for short) ontology-based image annotation tool was developed to address these issues and facilitate construction of Web image repositories with described semantics and emotion.

---

[†] – Corresponding author

The WNtags allows manual and collaborative image tagging with weighted sets of concepts. This tagging process is focused on high-level semantic and emotion content. The current version of the WNtags uses only WordNet as the tag glossary but its system architecture can accommodate concepts from other truly formal ontologies as image semantics descriptors. In this respect WordNet [1] is a useful knowledge base because of two reasons: 1) it defines a very large and supervised tagging glossary, and 2) the tags are structured within knowledge taxonomy. However, WordNet is not an ideal knowledge base for labeling and retrieval of images which will be discussed further in the article.

Apart from providing a user friendly Web interface for image annotation, the WNtags is also a querying tool that generates ranked sets of emotionally annotated real-life images. The image retrieval and raking algorithms are both founded on graph-based semantic similarity between search query and annotations of all stored images with the addition of weighting constants which modify the relevance of individual tags.

The WNtags is different from other similar tools for manual image annotation (for example see [2] and [3]) in several aspects. First, the WNtags uses International Affective Picture System (IAPS) dataset with semantically and emotionally described photographs that induce positive, neutral or negative affective reactions in subjects [4]. Since IAPS's free-text keywords are inadequate for an efficient image annotation [5], the WNtags with its optimized annotation glossary can also be interpreted as a tool that improves functionality and usability of this and other similar databases. Second, the WNtags' modular construction enables experimentation with different algorithms for image retrieval. Third, the WNtags uses WordNet for description of image content and in the future envisages applying Suggested Upper Merged Ontology (SUMO) [6] to formally schematize high-level image concepts. Finally, as mentioned before the WNtags uses weighted measures and heuristic function for image relevance estimation to more accurately rank semantically important concepts and to reconcile different measures of observer's variability.

The remainder of this paper is organized as follows. Section 2 offers an overview of emotionally annotated databases, their current status and explains how high-level image features can be semantically and emotionally annotated. Section 3 discusses the WNtags as the collaborative web-based research tool in the area of image annotation, knowledge representation and document retrieval and describes the tool's architecture. Section 4 specifies the data model for formal annotation of high-level image semantics. In Section 5 an image retrieval experiment is described and the obtained results are elaborated. Finally, Section 6 provides insight into future work.

## 2 Description and retrieval of emotionally annotated pictures: state of the art

Emotionally annotated databases are libraries of multimedia documents with annotated semantic and affective content that elicit emotional responses in exposed human subjects. They are successfully used for a wide variety of research in psychology, neuroscience and cognitive sciences for research in attention and emotion processing. The aforementioned International Affective Picture System (IAPS) and the International Affective Digital Sounds system (IADS) [7] are two of the most cited tools in the area of affective stimulation, although a number of other similar databases exist (for example [8] and [9]).

All multimedia documents indexed in emotionally annotated databases have specified semantic and emotional content. Two predominant theories used to describe emotion are the discrete category model and the dimensional model of affect [10] [11]. All databases have been characterized according to at least one of these models. The dimensional theories of emotion propose that affective meaning can be well characterized by a small number of dimensions. Dimensions are chosen based on their ability to statistically characterize subjective emotional ratings with the least number of dimensions possible [12]. These dimensions generally include one bipolar or two unipolar dimensions that represent positivity and negativity, and have been labeled in various ways, such as valence or pleasure. Also usually included is a dimension that captures intensity, arousal, or energy level. In contrast to the dimensional theories, categorical theories of emotion claim that the dimensional approach, particularly when using only two or three dimensions, does not accurately reflect the actual biological neural systems underlying emotional responses. Instead, supporters of these theories propose that there are a number of emotions that are universal across cultures and have an evolutionary and biological basis [13]. Most discrete models have in common at least the five primary emotions: happiness, sadness, anger, fear and disgust.

In currently available emotionally annotated databases a single stimulus is semantically described with a single tag from an unsupervised glossary. This semantic corpus does not have an internal knowledge framework defining semantic relations between different concepts nor prohibits usage of different keywords designating the same concept. For example, if a picture portrays an attack dog it may be tagged as „dog", „attack", „attackdog", „attack_dog" etc. Synonyms like „canine" or „hound" would be interpreted as different tags. Since this corpora has no semantic similarity measures there are no criterions to estimate relatedness between concepts. Therefore in these databases it is impossible to establish that „dog" and „cat" are more closely related than „dog" and

„Space Shuttle". This represents a huge defect in the retrieval process, because a user's search query has to lexically match the keywords stored in the database. No higher and more meaningful interpretation of either the query or annotating tags is possible. All these limitations in practice dictate that a person working with a contemporary emotionally annotated database has to be a well-trained expert in all its keywords, semantics and the annotated pictures. This type of skills is hard to acquire; even if one would possess them, it would be applicable only to a single domain, since the next database uses different keywords with different semantics.

The inadequate semantic descriptors result in three negative effects which impair retrieval: 1) low recall, 2) low precision and high recall or 3) vocabulary mismatch. The low recall results in a small set of retrieved pictures although the database contains a larger number of positive results. In the second effect the retrieval results in a high number of documents which do not semantically or affectively match the search query. The vocabulary mismatch represents an error in interpretation of the query. As a result the retrieval procedure returns documents with different semantics, not stated in or implied by the query. To overcome these problems a number of different approaches have been proposed in the literature [14].

The simplest and the most often used solution in image repositories on the Web are tag clouds with supervised vocabulary (for example see [15]). Tag or word clouds are weighted sets of keywords depicting high-level document features in a multimedia database or repository. They can be selected manually or automatically by an image recognition algorithm and adjoined to a picture. In this scheme a single object is described with at least one tag and its weight, or importance, is directly proportional to the number of objects to which the tag has been applied. Since documents may share the same keywords or tags it is possible to establish a semantic associations across different media based on the shared keywords. Given a compact glossary with human-understandable tags and a minimal number of different tags for the same concept (preferably without any semantic overlapping in the glossary), such approach may be good enough for an Internet image repository. However, in order to eliminate subjectivity of picture descriptors and given the volume of information to deal with it is necessary to use interoperable description schemes and provide for an automatic process of extracting semantic descriptions.

The second often employed solution to the retrieval problem is the application of WordNet for definition of tag glossary and concept taxonomy. WordNet is a well-known lexical database of English language which may also be viewed upon as an informal lexical ontology. It is readily available, simple to use and – most importantly – contains over 100,000 concepts, i.e. synsets. WordNet's large glossary encompasses almost any label a user may query and is ideal for informal description of high-level terms. Apart from the large glossary that in practice encompasses nearly all concepts that are likely to be used for image annotation, the other main advantage of

WordNet are numerous available algorithms for semantic relatedness and word-sense disambiguation [16]. Relying on WordNet makes it easier to calculate distances between concepts in queries and annotations, even if expressed as full sentences. Indeed, much work has been put into the development of optimal algorithms that produce results as close as possible to a human expert. Neither approach has been shown to be superior in all application domains but the work on this problem is still ongoing. Nevertheless, the main drawback of WordNet in image annotation is that linguistic tokens lack the formality of ontological concepts. Also, WordNet's taxonomy cannot offer enough expressivity that is required for a rich semantic representation of picture content. In spite of these shortcomings, WordNet is still far superior in image annotation than the existing keywords used in emotionally annotated databases which are inconsistent, ambiguous and non-contiguous [5][17].

Optimally, an image annotation procedure should not restrict the selection of text labels for describing images. This is contradictory to the need to reach an unambiguous understanding about the meaning of the tags and the picture content. Previous work on tagging IAPS pictures with WordNet concepts [18] showed that most users prefer to provide short and basic-level object labels (e.g. "child", "car", "person", "airplane"). This is not enough to establish a rich tag glossary which would enable quality image retrieval. At the same time, there can be a large variance of terms describing the same object category adding to the semantic ambiguity. Taken together, these make analysis and retrieval of the object labels difficult, since the system has to be aware of label synonyms, subsumed labels, member labels, and distinguish between object identity, events executed by objects, and attributes of objects and events.

## 3 WNtags collaborative web-based tool for manual image annotation

The WNtags is a Web-based tool for collaborative manual image annotation, knowledge representation and image retrieval. The tool's architecture (Fig. 1.) consists of three main modules: for image retrieval, for image annotation and for storage. These components are interdependent and consist of several subsystems. Image retrieval and image annotation modules each have their own graphical user interfaces (GUIs) to be used by the WNtags users and expert image annotators, respectively. The image retrieval module also contains result ranking subsystem which sorts the retrieved results according to their relevance to the query. Image annotation module and result ranking subsystem communicate directly with the WordNet knowledge base which in turn uses semantic distance calculator to obtain relatedness or distance measures between pairs of WordNet concepts. The calculator may use different algorithms to establish semantic distances, but in the current version the WNtags relies on WordNet::Similarity [19]. If similarity has to be calculated between a pair of concepts, the calculator consults its database and gets the stored value. This database is huge, since it must store every combination tuple, but is also fast, since its search algorithm is of a linear complexity. Decision for this type of architecture was motivated purely by the execution speed and system modularity. Storage module is based on a Relation Database Management System (RDBMS) and Structured Query Language (SQL) to insert, update and select image objects and their adjoined data. The storage module contains images, image tags and image tag weights. Each image tag is represented with text and its WordNet synset identification number, and weighted with a decimal number indicating its perceived relevance as an image descriptor. This data model is further explained in the next section.

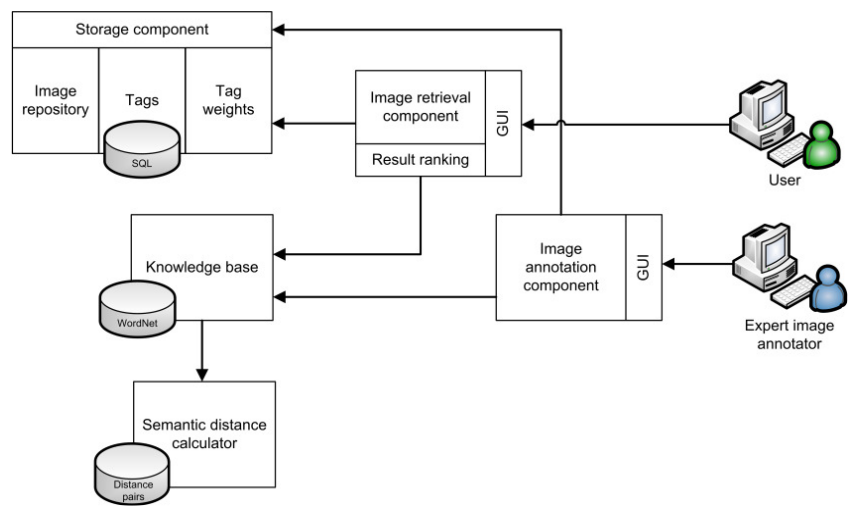

**Fig. 1.** System architecture overview.

The WNtags development was motivated by a number of objectives which can be aligned with two broad and complementary areas within the computer science and artificial intelligence: knowledge representation and image retrieval.

The most important goal of this project was to develop a collaborative online tool that could be modified and customized with minimal effort. In doing this, at least in the first version of the WNtags, the system requirements were confined to representation of high-level semantics in static images. Compromises had to be made since we wanted semantics annotations to be formal, but at the same close to a natural language. With the aid of a reasoning engine, the semantics annotations should be able to yield new knowledge about the image content. This is by no means an easy task, so we decided to use WordNet glossary to tag objects, events and affect in images. Our aim was to get an exhaustive set of relevant high-level semantics for each image in the database described with tags from a controlled glossary with graph-like structure. Such tag vocabulary can describe minimum necessary functional relationships between tags and discriminate among different meanings of the same lexical unit (i.e.

word). Furthermore, we wanted all tags to be weighted to indicate their importance in aggregation of image content.

Representation of picture segments and objects is currently not in our focus as with some other previously mentioned image annotation tools [2] [3]. However, we plan to scale out the tool to be able to annotate and store lower level semantics as well. In the future we also want to add support for other multimedia formats like sounds or video-clips, but foremost for specific domains of static images such as facial expressions and especially affectively annotated images. Our goal is to optimally reuse the existing code base and gradually build up the application's features.

The secondary goal behind the WNtags project was experimentation in information retrieval using the described weighted high-level semantics from a controlled graph glossary. Instead of using statistical methods and bag-of-words concept in document indexing and retrieval, we would like to see images precisely annotated with the least number of semantically meaningful tags. With formal logical systems based on Description Logic (DL) [20] these annotation may constitute a knowledge base of an intelligent system. Through the process of reasoning, the annotations could infer new knowledge about image content, effectively expanding the initial set of image tags. In this scheme, a search query given by a WNtags user becomes a search strategy goal in which the system must find the most closely matched images, fetch them from the database and sort the results by relevance. Procedures of user feedback and information refinement are also possible.

Finally, the last but perhaps one the most important incitements, the WNtags could also be interpreted as an experimental step towards improvement of emotionally annotated databases. As mentioned before, their description schemes of multimedia semantics are currently quite rudimentary and limiting, making image retrieval difficult and operator exhaustive [5]. Enhancement of multimedia content retrieval in these databases could substantially extend their use and promote further applications in cognitive sciences, psychology, psychiatry and neuroscience among others areas.

## 4 Formal annotation of high-level semantics

As explained previously, only a knowledge base using a large upper ontology and extended with appropriate domain ontologies can encapsulate semantics present in an image repository [21]. Naturally, domain ontologies would be selected depending on the repository's purpose and integrated with the upper ontology. Creating such image repository with large knowledge base and filling it with appropriately annotated images is an immense task. However, a compromise can be made with selection of WordNet as the tagging glossary.

In the WNtags, images are manually annotated with individual senses $s_1, s_2, \ldots, s_n$ of WordNet synsets $syn_1, syn_2, \ldots, syn_m$. Each image is tagged with at least three senses by a group of two or more individuals. In total 12 individuals manually described high-level semantic content of the images in the WNtags database. Every sense $s_i$ is weighted in a range $w_i \in [0,1]$ according to its perceived relevance. Apart from weighted semantic labels, each image $img_i$ retained its original semantic description from IAPS $iaps_i$. IAPS keyword vocabulary $V^{IAPS}$ is unsupervised and each IAPS picture is described with a single free text keyword. Effectively, $iaps_i$ is one keyword from this corpus. IAPS uses the aforementioned dimensional (i.e. circumplex) model of affect to annotate media emotion values [4][10][11]. In a nutshell, this model numerically describes emotion in emotionally-annotated picture as a tuple containing dimensions of valence or pleasure (denoted *val*), excitation or arousal (*ar*) and dominance (*dom*). All dimensions have normalized decimal values $val \in [1,9], ar \in [1,9], dom \in [1,9]$.

Every image $img_i$ in the WNtags repository retained its original IAPS emotional tuple value $emo_i$ with values $val_i$, $ar_i$ and $dom_i$. If $sem_i$ is set of all WordNet senses for $img_i$, $\hat{sem}_i$ set of weighted senses $sem_i$, and $iaps_i$ is one keyword from IAPS vocabulary then the cumulative semantic description $desc_i$ for one image $img_i$ stored in the WNtags image repository from becomes a tuple

$$desc_i = \{\hat{sem}_i, emo_i, iaps_i\} \tag{1}$$

where

$$sem_i = (s_{i_1}, s_{i_2}, \ldots, s_{i_n})$$

$$\hat{sem}_i = (w_{i_1} s_{i_1}, w_{i_2} s_{i_2}, \ldots, w_{i_n} s_{i_n})$$

$$emo_i = (val_i, ar_i, dom_i)$$

$$iaps_i \in V_i^{IAPS}$$

(2)

For every sense $s_i$ the WNtags considers WordNet's built-in semantic relations hypernymy, hyponymy, holonymy and meronymy within a preset node distance *d*. The WNtags also considers synonyms (coordinate terms) of $s_i$. Therefore for every sense $s_k$ in WordNet database and every sense $s_i$ in $desc_i$ broadens image description with all senses of all its neighboring synsets

$$sem_i = (s_{i_1}, s_{i_2}, \ldots, s_{i_n}) \cup \left( \bigcup_{s_j = s_{i_1}, s_{i_2}, \ldots, s_{i_n}} \left| (s_j, s_k) \right| \leq d \right) \tag{3}$$

As an example, IAPS annotated image (7175.jpg) with weighted labels and interaction with currently implemented WordNet knowledge base is shown in Fig. 2.

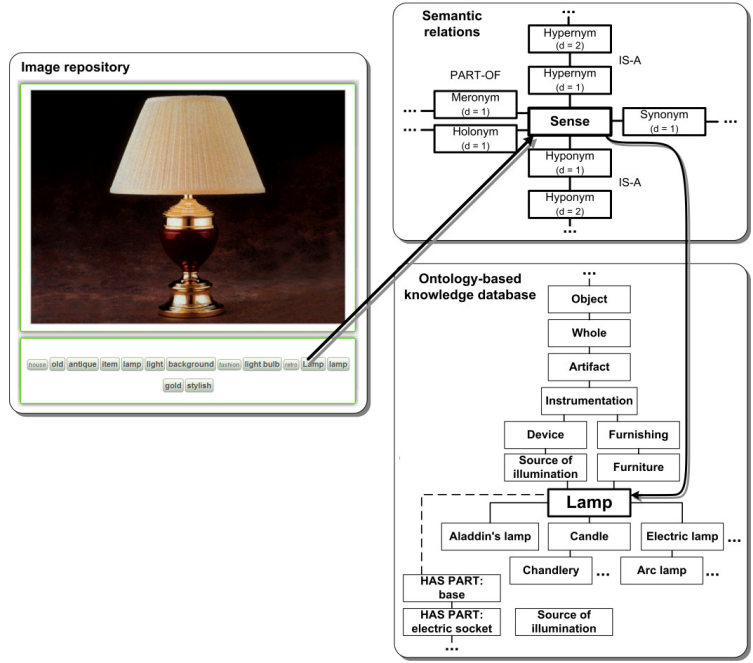

**Fig. 2.** IAPS image 7175.jpg stored in the WNtags database. The image is manually annotated with weighted concepts from a supervised glossary and described with WordNet lexical ontology. Each concept (e.g. "Lamp") has semantic neighborhood defined with semantic relations and node distance $d = 1, 2, 3, \ldots$

All indexed images are manually and collaboratively annotated. Personal bias of annotators will certainly result in different weights assigned to the image labels. Therefore, if there are $k$ weight ratings for $s_i$ they are averaged out $\bar{w}_i = \frac{1}{k} \sum_{j=1}^{k} w_{ij}$. Almost all users agreed on sets of 3-5 distinct labels per single image while finding additional descriptions in many cases proved difficult especially for semantically simple images.

In order to prove the validity of the WNtags concept we have randomly selected 100 emotionally annotated images with different object categories from IAPS database and tagged them using 956 different WordNet synsets. On average a single image is annotated with median of 20 WordNet tags (mean = 20.56436, sd = 2.76917, min=13, max=28) which were weighted according to their subjective importance in a particular image. The results are discussed in Section 5.1.

## 5  Image retrieval

The described WNtags annotating configuration allows multimodal image retrieval by three different information dimensions: affect, original free-text keywords and sets of synonyms from the WordNet lexical ontology. The WNtags has modular architecture and in the current version, to save processing time, semantic similarity is loaded from the freely available WordNet::Similarity dataset with prepared similarity and relatedness pairs [19].

The search algorithm uses all senses of all collocations (i.e. non-permutable combinations) of query concepts. Some more complex multiword collocations may be found as specific senses in WordNet. All such detected query senses $qs_i$ in query $q_k$ are individually aligned with annotating senses $s_i$ for all images $img_l$, $l = 1, 2, ..., N_{img}$ where $N_{img}$ is the total number of images stored in the repository. Semantic distance between each pair is calculated together with subjective importance $\bar{w}_j$ of each $s_j$ in image $img_l$. Search goal function is defined as

$$\max \sum_{\forall qs_i \in q_k} \sum_{\forall s_j \in img_l} \bar{w}_j sim(qs_i, s_j) \quad (1)$$

An exhaustive search is executed and the retrieved results are sorted from the highest to the lowest relevance with possible values in range. The final results are rendered and displayed to the user.

### 5.1  Results

To test the retrieval performance, we performed semantic searches. All the queries we submitted during the testing phase consisted of one WordNet concept randomly selected within $d = 30$ node distance between the nearest image tag, for example (as in Fig. 3.): "aircraft", "car", "boat", "helicopter", "road", "bus", etc. . The query tags that were used were chosen among those which were stored in the database during image annotation, thus ensuring we would get at least one relevant result.

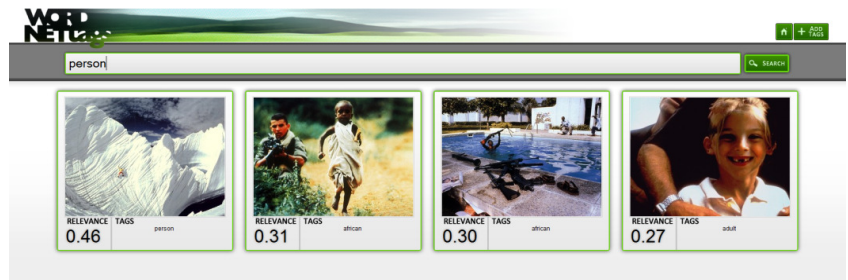

**Fig. 3.** Retrieved results from the WNtags repository after querying it with concept "Person".

After *N* = 40 one word queries, the average precision was 68.93 % and average number of retrieved true positive images (i.e. TP count) 6.15. As expected, different queries resulted in different precision, result and TP count. Multi word queries (with more than one WordNet concept) were not used in this experiment. Some queries yielded only one IAPS image, while others produced over 15 images ranked according to their semantic similarity with the querying concepts. Precision was calculated for each sequence number in the result set and TP count was normalized accordingly. The highest TP count (*TP images* = 20) was for the query "animal" because of the repository content and the fact that the most images were already manually tagged with this concept. As can be seen in Fig. 4., precision and TP count decrease as the number of retrieved images increases, as expected. The first result has precision 84.21 %, the second 81.58 %, the third 65.78 % and so on. Currently, the number of indexed images (*N* = 100) is too low for a thorough evaluation of retrieval performance. However, some indicative conclusions can be drawn from these queries. The aggregated and averaged results for all *N* = 40 one word queries are shown in Fig. 4. with TP count on x-axis and value range of precision and TP count on y-axis. Precision of all queries is indicated with a blue dotted line, while TP count is represented with a solid red line.

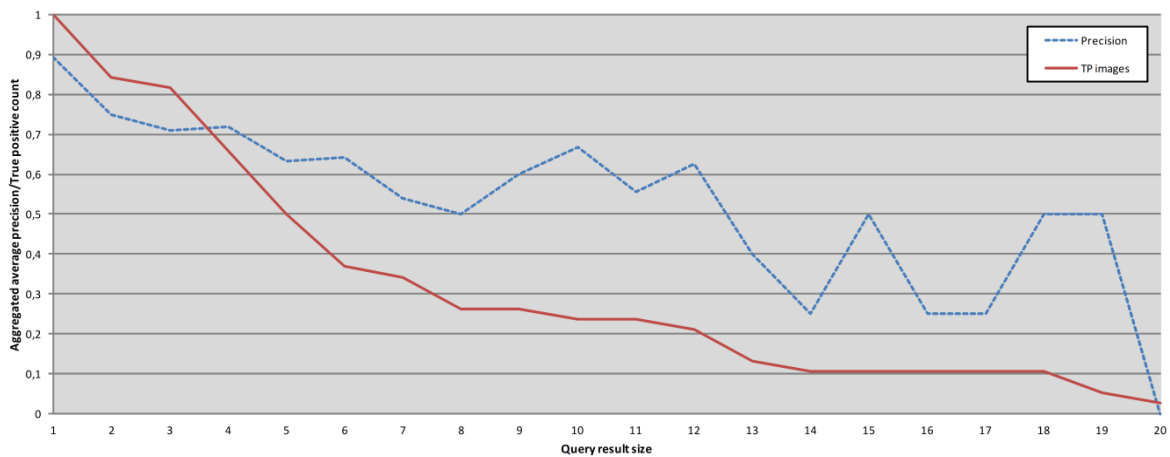

**Fig. 4.** Average retrieval performance after *N* = 40 queries with unit concepts. Average precision is 68.93 % and average number of TP images 6.15. The highest precision is 84.21 % for the first result in a retrieved sequence.

**5.2 Discussion**

Overall, we found images to be tagged quite exhaustively, which is good for establishing the ground truth in image content, but can be an obstacle for retrieval with semantically similar concepts leading to unrealistically high precision. As can be seen in Fig.3, precision oscillates as the number of results increases, which is a conse-quence of this over-tagging problem. This problem may be easily resolved if only a randomized subset of tags is used in retrieval.

The second problem is how to define a realistic, useful and at the same time generic measure of semantic relatedness between concepts in ontology. A variety of information-based (corpus dependent) and node distance-based (information independent) similarity measures have different successes in diverse application domains. Many such measures exist and finding the most suitable semantic similarity measure is an ongoing knowledge representation problem. Another related problem is objectivity of tag weight coefficients values. Although potentially a separate source of information valuable for image retrieval, weight values are a product of individual, and subsequently, potentially biased criterion. Tags with unequally distributed weights will produce erroneous relatedness with query tag, especially when coupled with suboptimal semantic distance metrics. However, *kappa* statistic may be used to indicate weights that have inadequate inter-annotator agreement. Tags with such weights should be re-annotated or outlier values left out and weights recalculated. Either way, tag weights have to be consistent across the whole image repository or they will deteriorate the semantic relatedness function. The best way to do this is to establish a clear and formal set of rules for calculating the weights so the influence of personal bias is annulled.

Third, the structure of WordNet's semantic network may be contrary to intuition and too complex [22]. Some of the top-level WordNet concepts such as "Artifact" or "Whole" may be useful for construction of a well-indented taxonomy, but they are also very much semantically ambiguous or redundant to an average person. Intuitively, it seems that a pruned down and lightweight version of WordNet, with new semantic relations such as "Relates-To", "Similar-To" or "Occurs-With", may be better as a knowledge taxonomy for image labeling than the complete WordNet, albeit without relations that may be lexically ambiguous but could provide fresh and important information content.

WordNet is a huge benefit, but at the same time also a burden. Expressivity of the WNtags image annotations is adequate to support content rich and diverse image extraction, but qualitatively the lack of formalism and complex functional relationships between concepts proved limiting and led to ambiguities with different concept senses. For example, a single WordNet concept "snake" has a synonyms "serpent". An average WNtags user expects that semantic distance between all synonyms and another different concept to be the same. However, this is not the case with node distance metrics based on WordNet senses; semantic neighborhoods in WordNet semantic graph around synonyms are not identical. This leads to different image retrieval results when querying these concepts, which may be confusing to the user.

Because of this reason it is unnecessary to observe a wide semantic neighborhood around the search query. Distance of up to 5 or 10 nodes is usually enough to pick up the majority of concepts related to a query. Larger distances will increase recall, but substantially decrease accuracy and precision. However, an adaptable algorithm may be envisioned to find a good balance between these two opposing parameters. Such algorithm could start with a small semantic neighborhood around the query and then iteratively increase the distance or include different semantic relations until enough related concepts are found or some other stopping criterion is met. Further research is required to validate this idea.

As mentioned before, it is obvious that only lexically rich knowledge base is not enough for high quality annotation and retrieval. Ontology with a sufficient number of annotating concepts representing objects, events and their various functional relations should allow more accurate image retrieval but also retain a rich annotating glossary which is desirable from the user's perspective.

## 6 Future Work

In the future we would like to retain existing WordNet image labels and align them with concepts from the SUMO ontology because only a large ontology deprived of lexical dependencies is able to formally represent the ground truth meaning of image content. This can be done efficiently with SUMO to WordNet lexicon mappings [23]. The mappings enrich WordNet knowledge base by tagging each synset with the corresponding SUMO concept. These mappings allow for an automated transfer of natural language words into SUMO terms, using WordNet synsets as an intermediate layer.

Another WNtags feature which requires further work is the semantic similarity calculator. To improve speed and reduce unnecessary overhead, the calculator was implemented as a large look-up table storing tuples $\{c_1, c_2, r(c_1, c_2)\}$ of two concepts $c_1$, $c_2$ and their relatedness value $r(c_1, c_2) \in [0,1]$. Similarity was *a priori* calculated for all different pairs of concepts that can be used for annotation and retrieval. However, a lot of false positive (i.e. Type I errors) retrieval results are a consequence of inaccurate similarity estimates between query and image concepts. Obviously, the estimates are not optimal or analogous to a human perception, and have to be amended. To find out the best algorithm we plan to modularly add different semantic similarity algorithms for information retrieval including information-based metrics.

It also would be beneficial to explore new modalities of indexing additional multimedia formats apart from pictures considering specific affective-related domains of static images such as facial expressions and video stimuli. A promising avenue for the experimental research would be the creation of a truly multimodal annotated database containing a fusion of different emotionally annotated formats including pictures, voices, natural and manmade sounds, music, powerful real-life video clips, etc. Such tool would be extremely advantageous for a wide range of interdisciplinary research that requires a large amount of different media with specific semantics, emotion and context, and fast, voluminous and accurate retrieval on the other hand.

# 7 Conclusion

A new model for description of high-level semantic features with WordNet concepts in emotion images has been presented in the article. Also, an overview of currently available emotionally annotated databases has been provided, with a clear outlook for their optimization. A rich formal description of object, events, scenes, context, emotion and other relevant media indexing information, combined with an efficient inference engine, is required to overcome present deficiencies in tagging of emotionally annotated images.

The WNtags tool was created to test the proposed assumptions. This is web-based image annotation and retrieval tool that may be used to manually index high-level semantic concepts and emotion in images, store them in a relational database, and query such image repository to obtain ranked sets of emotionally annotated images. Owing to the tool's friendly user interface, new images can easily be added to the repository and labeled with WordNet concepts. The WNtags performs image retrieval using WordNet's lexical ontology topology, node distance metrics and collaboratively weighted tags. The results are sorted relative to semantic similarity between concepts in image annotations and search queries, and their importance in description of each individual image.

Future implementation of SUMO upper ontology, WordNet to SUMO mappings and different semantic similarity algorithms are expected to significantly improve the retrieval. Import of further emotionally annotated images from different sources will also provide new experimental insight into the optimization of emotionally annotated databases.

# Acknowledgements

This research has been partially supported by the Ministry of Science, Education and Sports of the Republic of Croatia, grant no. 036-0000000-2029. The WNtags tool was developed by a student team: Anton Grbin (project leader), Aleksandar Dukovski, Anton Grbin, Jerko Jurin, Dino Milačić, Matija Stepanić, Hrvoje Šimić and

Dominik Trupčević as part of their Programming Project course at the University of Zagreb, Faculty of Electrical Engineering and Computing. The WNtags tool can be accessed by contacting the first author of the paper.